\documentclass[trackchanges]{aastex631}

\usepackage{savesym}
\savesymbol{tablenum}
\usepackage{siunitx}

\restoresymbol{SIX}{tablenum}

\newcommand{\alf}{Alfv$\acute{\mathrm{e}}$n}

\submitjournal{ApJ}

\begin{document}

\title{A Machine Learning Approach to Understanding the Physical Properties of Magnetic Flux Ropes in the Solar Wind at 1 AU}

\correspondingauthor{Hameedullah Farooki}
\email{haf5@njit.edu}

\author[0000-0001-7952-8032]{Hameedullah Farooki}
\affiliation{Institute for Space Weather Sciences,
New Jersey Institute of Technology,
University Heights, Newark, NJ, USA}
\affiliation{Department of Computer Science,
New Jersey Institute of Technology, University Heights, Newark, NJ, USA}
\affiliation{Center for Solar-Terrestrial Research,
New Jersey Institute of Technology,
University Heights, Newark, NJ, USA}

\author[0000-0003-0792-2270]{Yasser Abduallah}
\affiliation{Institute for Space Weather Sciences,
New Jersey Institute of Technology,
University Heights, Newark, NJ, USA}
\affiliation{Department of Computer Science,
New Jersey Institute of Technology, University Heights, Newark, NJ, USA}

\author[0000-0002-8032-7833]{Sung-Jun Noh}
\affiliation{Center for Solar-Terrestrial Research,
New Jersey Institute of Technology,
University Heights, Newark, NJ, USA}
\affiliation{Now at Los Alamos National Laboratory, Los Alamos, NM}

\author[0000-0002-6350-405X]{Hyomin Kim}
\affiliation{Institute for Space Weather Sciences,
New Jersey Institute of Technology,
University Heights, Newark, NJ, USA}
\affiliation{Center for Solar-Terrestrial Research,
New Jersey Institute of Technology,
University Heights, Newark, NJ, USA}

\author{George Bizos}
\affiliation{Department of Computer Science,
New Jersey Institute of Technology, University Heights, Newark, NJ, USA}

\author[0000-0002-0815-9855]{Youra Shin}
\affiliation{Center for Solar-Terrestrial Research,
New Jersey Institute of Technology,
University Heights, Newark, NJ, USA}

\author[0000-0002-2486-1097]{Jason T. L. Wang}
\affiliation{Institute for Space Weather Sciences,
New Jersey Institute of Technology,
University Heights, Newark, NJ, USA}
\affiliation{Department of Computer Science,
New Jersey Institute of Technology, University Heights, Newark, NJ, USA}

\author[0000-0002-5233-565X]{Haimin Wang}
\affiliation{Institute for Space Weather Sciences,
New Jersey Institute of Technology,
University Heights, Newark, NJ, USA}
\affiliation{Center for Solar-Terrestrial Research,
New Jersey Institute of Technology,
University Heights, Newark, NJ, USA}
\affiliation{Big Bear Solar Observatory, New Jersey Institute of Technology, 40386 North Shore Lane, Big Bear City, CA 92314, USA}



\begin{abstract}

Interplanetary magnetic flux ropes (MFRs) are commonly observed structures in the solar wind, categorized as magnetic clouds (MCs) and small-scale MFRs (SMFRs) depending on whether they are associated with coronal mass ejections.
We apply machine learning to systematically compare SMFRs, MCs, and ambient solar wind plasma properties. We construct a dataset of 3-minute averaged sequential data points
of the solar wind's instantaneous bulk fluid plasma properties using about twenty years of measurements from \emph{Wind}.
We label samples by the presence and type of MFRs containing them
using a catalog based on Grad-Shafranov (GS) automated detection for SMFRs and NASA’s catalog for MCs
(with samples in neither labeled non-MFRs).
We apply the random forest machine learning algorithm to find which categories
can be more easily distinguished and by what features.
MCs were distinguished from non-MFRs with an AUC of 94\%
and SMFRs with an AUC of 89\% and had distinctive plasma properties.
In contrast, while SMFRs were distinguished from non-MFRs with an AUC of 86\%,
this appears to rely solely on the $\langle B \rangle > \SI{5}{\nano\tesla}$ threshold applied by the GS catalog.
The results indicate that SMFRs have virtually the same plasma properties as the
ambient solar wind, unlike the distinct plasma regimes of MCs.
We interpret our findings as additional evidence that most SMFRs at 1 au are generated
within the solar wind, and furthermore, suggesting that they should be considered
a salient feature of the solar wind's magnetic structure rather than transient events.

\end{abstract}



\section{Introduction} \label{sec:introduction}

Magnetic flux ropes (MFRs) are structures of plasma in space characterized by their spiraling magnetic field lines
observed throughout the heliosphere with a wide range of shapes and sizes.
They are believed to be the core structure of many phenomena in the atmosphere of the Sun such as
prominences/filaments \citep{gibson2018prominencereview,liu2020fluxropecorona}.
As observed in the solar wind near Earth,
MFRs are usually classified as either magnetic clouds (MCs) or small-scale magnetic flux ropes (SMFRs)
\citep{hu2018automated}.
Both MCs and SMFRs have been detected in all locations of the heliosphere
at which the necessary measurements are available,
from near the Sun \citep{zhao2020identification,yuchen2020_gs_psp_first_two},
to near the Earth \citep{moldwin2000smfr,hu2018automated,nieves2018icmecatalog},
to above the ecliptic plane \citep{chen2019analysis}
all the way out to 8 au \citep{chen2020effects}.
Although they have similar magnetic structures (hence they are both referred to as MFRs) and are both observed in the solar wind near Earth, MCs and SMFRs are very different and may have different origins. MCs are larger and are the internal magnetic structure of interplanetary coronal mass ejections (ICMEs; coronal mass ejections traveling through the solar wind), whereas SMFRs are smaller, found scattered throughout the bulk of the solar wind, and have an uncertain origin (or multiple origins). MCs and SMFRs differ significantly in their plasma properties, e.g. MCs have very low temperature and plasma beta whereas SMFRs do not always have these properties.

MCs were originally reported with a strict definition by \citet{burlaga1981magnetic}, although the usage of the term has become more general over time \citep{nieves2018icmecatalog}.
Typically observed a few times a month with durations on the order of tens of hours at 1 AU,
they are widely accepted to approximately be the internal magnetic structure of many,
if not all, interplanetary coronal mass ejections (ICMEs) 
\citep{hu2014structures,chen2017physics,nieves2018icmecatalog}, which are coronal mass ejections departed from the sun 
propagating through interplanetary space \citep{howard2009interplanetary, kilpua2017coronal}).
ICMEs have a significant impact on space weather \citep{baker2004effects}
and thus there is much interest in their study.

In contrast, SMFRs occur frequently near Earth, with a few hundred observed monthly on average
and a strong solar cycle dependency
\citep{hu2018automated}.
Unlike MCs, the origin of SMFRs is not firmly established
\citep{borovsky2008flux,hu2018automated,chen2022small,sanchez2017observational,sanchez2017temporal,sanchez2019situ,rouillard2010intermittent,rouillard2011solar}.
They are also much smaller with a typical duration of under an hour at 1 AU
\citep{hu2019radial}.
Moreover, their scale sizes range multiple orders of magnitude
and their physical properties vary with size \citep{hu2018automated}.
Furthermore, detecting SMFRs is challenging as the observational signatures
are not as clear as those of ICMEs.
Understanding SMFRs is important due to their ubiquitous presence in the solar wind,
potential significance for larger-scale phenomena such as solar wind acceleration,
and their close relationship with MCs.
There has been increased interest in detecting and analyzing SMFRs in recent years,
particularly near the Sun using new data from the Parker Solar Probe
\citep{chen2022small}.

SMFRs were first presented by \citep{moldwin1995ulysses, moldwin2000smfr}.
They identified one SMFR from measurements of the \emph{Ulysses} spacecraft
and six SMFRs from two other spacecraft, \emph{Wind} and IMP 8.
They found that SMFRs were very similar to MCs but much smaller
and with some different plasma properties.
Multiple statistical studies were conducted in later years,
but they were typically limited to no more than a few hundred samples
\citep{hu2018automated}.
More recently, \citep{zheng2017automated} developed
a novel automated detection algorithm based on the Grad-Shafranov (GS)
reconstruction technique that could detect tens of thousands of SMFRs from
about twenty years of \emph{Wind} observations.
This event list was then formally published by \citep{hu2018automated}.
Since then, the GS-based automated detection methodology has been
extended to various spacecraft throughout the heliosphere
such as \emph{Ulysses} and the Parker Solar Probe
\citep{chen2019analysis,yuchen2020_gs_psp_first_two,chen2022small}.

Machine learning is a part of artificial intelligence with algorithms
for fitting models to data
that allow computers to learn patterns in the data
and produce predictions for new data
or provide insights about the current data
\citep{alpaydin2020introduction}. Machine learning has seen
increased use in a wide range of disciplines including space physics
(e.g.
\citet{abduallah2022predicting};
\citet{liu2020predicting};
\citet{vech2021novel};
\citet{zewdie2021data};
\citet{richardson2021seasonal};
\citet{roberts2020objectively};
\citet{raheem2021investigation}).
This is because machine learning is a versatile tool
useful when dealing with large amounts of data.
However, there have been limited attempts at applying machine learning
to studying MFRs in the solar wind.
Recently, there have been a few machine learning applications to MCs:
\citet{dos2020identifying} used a deep convolutional neural network to
detect signatures of simple MCs in ICMEs,
and \citet{narock2022identification} also used a deep convolutional network
in order to identify the axis of a known MC.
\citet{nguyen2019automatic} developed a deep learning method to 
detect ICMEs in general.
\citet{reiss2021bz} used machine learning algorithms
to predict the $B_z$ component of an ICME.
However, there is still much room for further application of machine learning to MCs.
Furthermore, to our knowledge, there have been no attempts to apply machine learning
to SMFRs.

In this paper, we apply a machine learning technique to better understand the importance of various physical properties during the presence (or lack thereof) of MCs and SMFRs. We do this by using a machine learning algorithm (known as random forests) to learn the probability distribution of a sample corresponding to various MFR categorizations (such as SMFR, MC, and non-MFR) based only on point-in-time physical properties. We use the detected events in the existing catalogs as training data. The purpose of our application of machine learning is not to detect MFRs, as point-in-time data is insufficient to properly detect MFRs (for example, detection algorithms use multiple points in time to detect a rotation in the magnetic field direction, not just looking at the direction at a single point in time). Furthermore, both MC and SMFR catalogs and detection algorithms are already available (see Section~\ref{sec:data}). Instead, our machine learning approach makes it possible to systematically compare the physical properties of the solar wind under various MFR conditions.

The contribution of our paper is different from previous studies. \citet{hu2018automated} compared the physical properties of SMFRs in the fast and slow solar wind and analyzed the overall statistics of SMFR properties. But a detailed and systematic comparison of SMFR properties to the background solar wind (when no SMFR is detected), as well as between SMFRs and MCs, is still needed and addressed in this paper. Although we use a machine learning algorithm to distinguish between MFRs and non-MFRs, our ultimate purpose is not to detect MFRs but to analyze the differences between MFR categorizations. Previous machine learning-based studies (e.g. \citet{camporeale2017classification} and \citet{li2020machine}) have introduced various methods to accurately classify solar wind regimes based on in-situ measurements. However, our study differs in its focus on MFR categorizations (especially SMFR) and its focus on the strength of the physical properties of the solar wind as probability estimators of the categorizations rather than the overall accuracy of the classifier.

The rest of this paper is structured as follows.
First, in Section~\ref{sec:data},
we describe in detail the input data and MFR event lists that we utilize.
We also provide a brief statistical analysis of the dataset.
Then, in Section~\ref{sec:methodology},
we describe the machine learning techniques that we use,
including the classification algorithm and feature selection procedure.
Next, in Section~\ref{sec:results},
we present the results of our experiment.
Finally, in Section~\ref{sec:discussion},
we discuss the results and prospectives for future work.
\section{Data} \label{sec:data}
\subsection{Input Features}

The fundamental physical aspects of the solar wind that we analyze
are the magnetic field and plasma properties.
Firstly, the magnetic field is the most relevant
due to its fundamental role in defining what an MFR is
(that is, a plasma structure with twisted magnetic field lines),
although the single timestamp measurements that we use
don't individually contain information about the spatial structure or temporal change.
The atomic composition of the solar wind plasma
is mostly of protons and then of alpha particles,
so we narrow down our focus to those two.
Since the data availability and count statistics for protons (which dominate the solar wind)
is higher than those of alpha particles, we only use proton parameters rather than
alpha particle parameters, other than the alpha/proton number ratio to indicate the
abundance of alpha particles.
Because of data gaps in the electron measurements, we opted to leave out
electron data for this study.

All input features come from observations from the \emph{Wind} spacecraft
provided by NASA's Space Physics Data Facility.
The time period of the data is from 1996 to 2016.
The magnetic field parameters are retrieved from the
Wind Magnetic Field Instrument (MFI)
dataset \citep{lepping1995wind} at 1 minute cadence,
and the plasma parameters are retrieved from the Soar Wind Experiment (SWE)
dataset \citep{ogilvie1995swe}.
We use vector parameters in the
Geocentric Solar Ecliptic (GSE) coordinate system.
The magnetic field parameters are the three components
of the magnetic field vector $\vec{B}$ ($B_x$, $B_y$, and $B_z$)
converted to altitude and azimuth ($B_{\theta}$ and $B_{\phi}$),
as well as the magnetic field strength $B$,
which is provided by the MFI data
averaged from higher cadence measurements.
The plasma parameters are the three components of the proton bulk flow velocity vector
$\vec{u_p}$ ($u_{p,x}$, $u_{p,y}$, and $u_{p,z}$)
converted to altitude and azimuth ($u_{p,\theta}$ and $u_{p,\phi}$),
proton bulk speed $u_p$,
proton number density $n_p$,
and the alpha/proton ratio $n_{\alpha}/n_p$.
Rather than including the absolute value of the proton temperature, we include the ratio between the observed and expected temperatures based on the proton speed \citep{richardson1995regions}.
Although we ignore temporal variations,
we included the directional components of the vector quantities
as there may be an overall tendency to different directions.
We converted to spherical coordinates to separate the magnitude from the direction.
Since the Wind/SWE data has an inconsistent cadence, all of the data is resampled
to 3-minute averages.
We excluded all data points in which any of the features had no data in the resampling bins.

In addition to the fundamental parameters,
relevant derived parameters are
added to the dataset in case they provide a stronger relationship.
Proton gas pressure is added, as it is important property of the solar wind plasma:
\begin{equation}
    P_{\mathrm{gas},p} = n_p k_B T_p
\end{equation}
where $k_B$ is the Boltzmann constant.
Another form of plasma pressure is the dynamic pressure
from the motion of the protons:
\begin{equation}
    P_{\mathrm{dyn},p} = \frac{1}{2} m_p n_p u_p^2
\end{equation}
where $m_p$ is proton mass.
The proton beta is a significant parameter because one of the most important
properties of MFRs, especially larger ones, is that they tend to have a low beta
\citep{klein1982interplanetary,hu2018automated}.
Using the gas pressure and magnetic field strength, the proton beta is calculated:
\begin{equation}
    \beta_p = \frac{P_{\mathrm{gas,p}}}{B^2/(2\mu_0)}
\end{equation}
Finally, two other plasma properties important in the solar wind
are {\alf} velocity $v_A$ and {\alf} Mach number $M_A = u_p/v_A$. 
$v_A$ plays an important role in magnetohydrodynamic fluctuations \citep{mullan2006solar}.
It is conceivable that differences in the density and magnetic field of MFRs
may result in significant differences in $v_A$.
This may also affect $M_A$,
which is an important indicator of the level of dominance of the magnetic field,
plays an important role in shocks \citep{sundberg2017dynamics},
and has an impact on the magnetosphere \citep{lavraud2013asymmetry}.
We include them both, calculating the {\alf} velocity with the equation:
\begin{equation}
    v_A = \frac{B}{\sqrt{2 m_p n_p \mu_0}}
\end{equation}

\subsection{Output Labels}

The labels are based on separate sources for MCs and SMFRs.
MC labels come from the ICME catalog developed by \citep{nieves2018icmecatalog},
which is available at (\url{https://wind.nasa.gov/ICME_catalog/ICME_catalog_viewer.php}).
This catalog is based on previous catalogs as well as additional visual inspection.
Most of the ICMEs have either well-defined MCs or similar structures \citep{nieves2018icmecatalog}, so we simply use the magnetic object start and end times provided by the catalog, regardless of the magnetic object type.

SMFRs come from
the event list on \url{https://fluxrope.info} developed by \citep{hu2018automated}
using the GS automated detection algorithm.
We use it because it is the most comprehensive list of SMFRs
from the \emph{Wind} spacecraft.
However, we found that most MCs were observed as a series of shorter SMFRs
due to the limited window sizes employed by \citet{hu2018automated}.
Therefore, we exclude
events overlapping with MCs (approximately 5\% of all SMFR events).

The SMFRs were identified by \citep{hu2018automated} by trying to identify flux rope structures using segments
of the time series observations of the solar wind from the \emph{Wind} spacecraft.
In contrast, this paper focuses on the properties of individual data points,
not structures measured over time, in order to focus on the physical properties
associated with MFRs. However, there is a caveat: In order to avoid detecting mere fluctuations
as SMFRs, \citet{hu2018automated} excluded SMFR candidates with
an average magnetic field strength below $\SI{5}{\nano\tesla}$.
As will become apparent later, this poses an issue for analyzing the physical properties of SMFRs.

We distinguish between SMFRs of durations under one hour (labeled short SMFR or SSMFR) and over one hour (labeled long SMFR or LSMFR). The reason for the distinction is that most SMFRs are under an hour in duration, so long SMFRs are unusual.
It is possible that LSMFRs and SSMFRs have different origins and impacts on the geospace system. 
Furthermore, \citet{hu2018automated} have noted physical differences between short SMFRs and long SMFRs. Finally, long SMFRs may be expected to have more in common with MCs than short SMFRs due to their more comparable size. Therefore, it is interesting to compare our results for short SMFRs and long SMFRs.

In order to generate the binary labels from the combined event lists, we
mark each timestamp contained within any MFR from either list
as positive and the rest as negative.
We also add categorical labels describing whether the MFR of a positive timestamp was an SMFR or an MC
and a duration label providing the MFR's duration. These labels are not directly used
by the machine learning models,
but rather they are used to generate the various
binary classification tasks described in Section~\ref{sec:methodology}
before being removed from the dataset.

All samples not included in either catalog were marked as non-MFRs. Therefore, a limitation of this study is that we assume that the event lists that we use are comprehensive and exclusive.
That is, we assume
that all of the events really were MFRs with correct and precise time boundaries and that all of the time periods with no events in
either list
contained no MFR whatsoever.
Nevertheless, the overall differences in statistical properties should be
reasonably reliable assuming that the event lists are mostly correct and complete.
As we will see, the main impact is that the difference between non-MFRs and SMFRs
is difficult to determine.

\begin{figure*}[ht!]
\includegraphics[width=.49\linewidth]{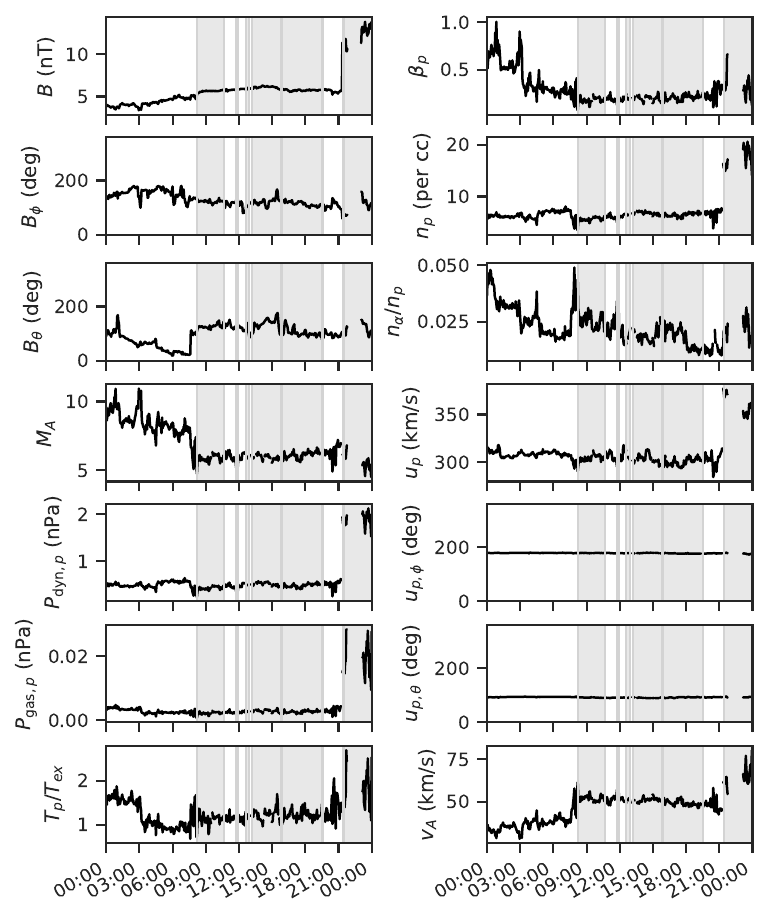}
\includegraphics[width=.49\linewidth]{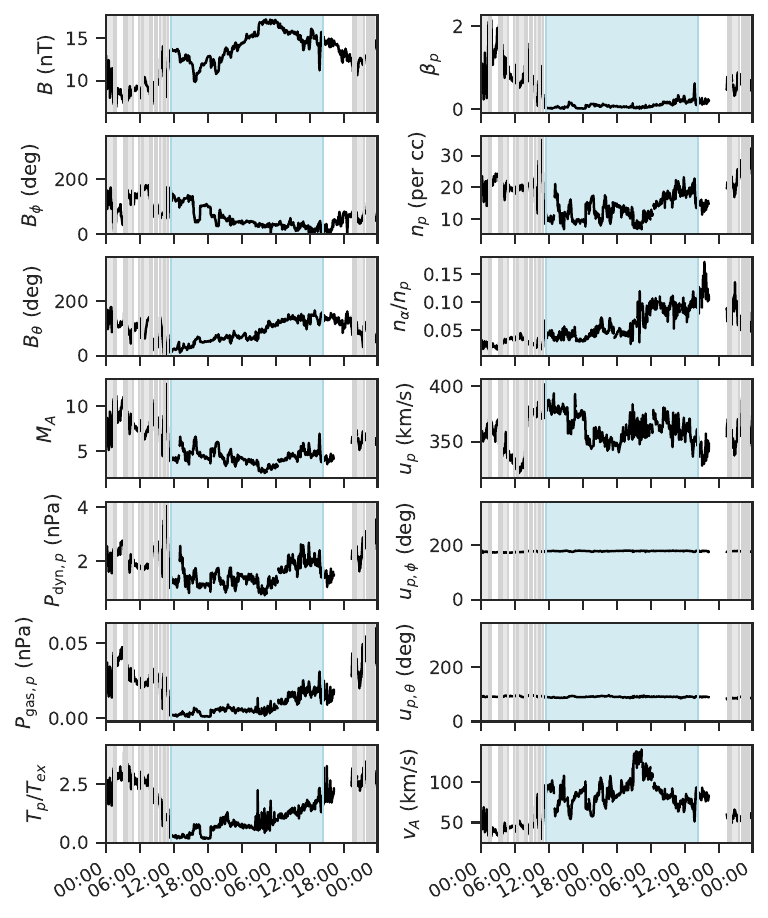}
\caption{(left) Data for the date 2016–01-18. Shaded regions are SMFR intervals. Data gaps in individual values are represented by breaks in the line plots.  (right) Data for dates 2016-01-19 and 2016-01-20. Same format as (left), except the region shaded blue contains an MC.}
\label{fig:visual}
\end{figure*}

Figure~\ref{fig:visual} (left) shows SMFRs on the date 2016-01-18. The shaded regions contain SMFRs according to the catalog. Each feature is plotted in its own subplot. Our dataset contains not entire regions, but individual data points, evenly spaced. For a given sample, there are 15 values corresponding to the 15 features and the aforementioned labels based on what region the sample’s timestamp falls under. Figure~\ref{fig:visual} (right) shows an MC that was observed between dates 2016-01-19 and 2016-01-20, in the blue-shaded region. Before and after the MC are SMFRs. In between the MC and SMFRs are unshaded regions, which were not included in either catalog. The samples falling under these regions are assumed to contain no MFR and are thus marked non-MFRs.

\subsection{Statistical Analysis}

The final dataset contains 15 features (described previously) and $2.8 \times 10^6$ samples.
Of these, $1.9 \times 10^6$ (69\%) contain no MFR,
$7.7 \times 10^5$ (31\%) contain any MFR,
$7.3 \times 10^5$ (26\% of the full dataset) contain an SMFR
(including $3.0 \times 10^5$ SSMFR and $4.3 \times 10^5$ LSMFR),
and $1.4 \times 10^5$ (5\% of the full dataset) contain an MC.
The total number of MFRs is 63807.
63461 of them are SMFRs (from the GS-based list, not overlapping with an MC)
and the remaining 346 are MCs (from the ICME list).
Of the SMFRs, 51,574 are SSMFRs and 11,887 are LSMFRs.

\begin{figure}
\centering
\includegraphics[scale=.9]{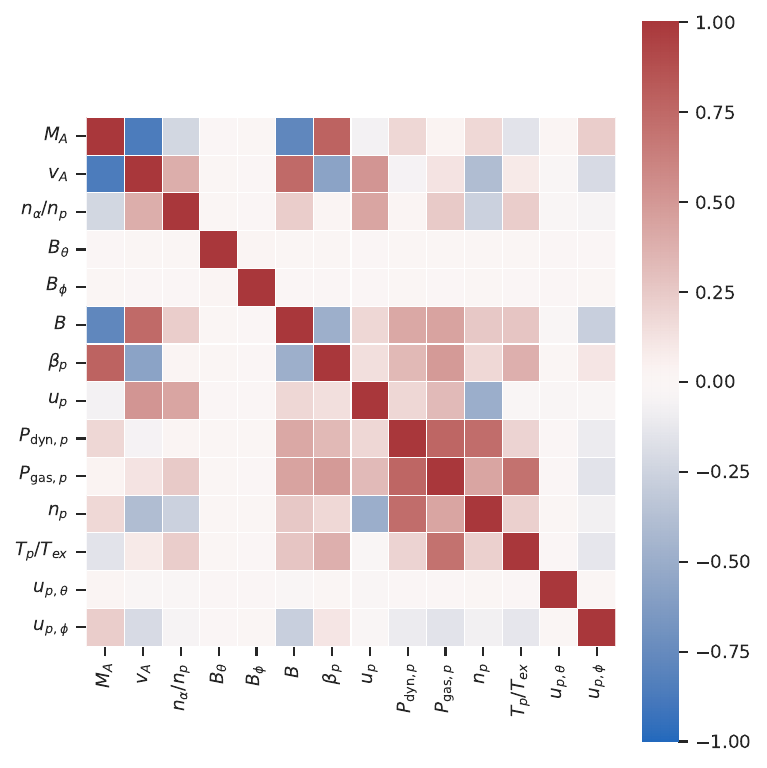}
\caption{Correlation heatmap. Shows Spearman correlation between features including raw measurements and derived parameters.
        Higher positive values indicate a co-increasing relationship,
        whereas lower negative values indicate an inverse relationship.
        Near-zero indicates a lack of any monotonously decreasing or increasing relationship.}
\label{fig:correlation}
\end{figure}

Considering correlations within the dataset is important both for understanding the data
and because correlations can impact feature importance scores (since a machine learning model could use
a threshold on either of two variables that have a strong correlation to get the same split of data samples).
Figure~\ref{fig:correlation} shows a heatmap of the pairwise correlation matrix of all the features.
The correlations are calculated using the Spearman correlation metric,
which is similar to the Pearson correlation metric
but shows any monotonically increasing (positive correlation)/decreasing (negative correlation)
relationship rather than only linear relationships.
The correlation matrix is symmetric, so it is the same across the diagonal. The diagonal contains all ones since
each feature is perfectly correlated with itself. While many feature pairs have almost no correlation,
there are some significant correlations. The strongest ones should be kept in mind when interpreting the results. 

\section{Methodology} \label{sec:methodology}
\subsection{Classification Tasks}

We define six binary classification tasks to be solved
by our model based on the data: SSMFR-NMFR, LSMFR-NMFR, SSMFR-LSMFR, MC-NMFR, MC-SSMFR, and MC-LSMFR. Each task consists of
estimating the likelihood of a sample's true label being positive
based on a subset of the data marked as positive merged with
another subset of the data marked as negative.
For example, SSMFR-NMFR has samples in SMFRs of duration under 1 hour as the positive
class and non-MFR samples as the negative class.
Unmarked labels are excluded from the dataset
for the corresponding task.
The different tasks extensively explore the MFR categorization of a point in time
and are defined so that every pair is compared.

\subsection{Algorithm}
We adopt the random forest classification algorithm \citep{RandomForestCART1984}
to generate our machine learning model.
This algorithm is well-suited because it tries random subsets of the features
on random subsets of the data to build an ensemble model.
The random sampling of features is to thoroughly explore different possible feature combinations
to give them a chance to demonstrate their predictive power and importance
with respect to the labels. On the other hand,
the random subsetting of the data, or bootstrapping,
is so that the trees are more diverse.
It is widely used in many domains
and has been used successfully for space weather science
in particular. For example,
\citep{liu2017rfsolarflare} used a random forest
in order to forecast solar flares in solar active regions.

In simple terms, the random forest algorithm learns by using a training set of
input data vectors and corresponding
output categorizations, or labels, to generate
an ensemble of randomized
decision trees that collectively vote
on the correct label of new input. The model
is trained as follows: Given a training set
consisting of $N$ vectors with $M$ features
$X = x_1, x_2, \cdots, x_{N-1}, x_N$
with labels $Y = y_1, y_2, \cdots, y_{N-1}, y_N$, generate $B$ binary trees $T_1, T_2, \cdots, T_{B-1}, T_B$
according to the following algorithm:
\begin{enumerate}
    \item Sample $N$ elements (with replacement)
        of the training
        set, denoted $(X_i, Y_i)$.
    \item Train binary decision tree
        $T_i$ on $(X_i, Y_i)$.
        At each split, a subset of $\sqrt{M}$
        features are selected to compare.
        The optimal feature for splitting is
        selected by maximizing the resulting
        information gain or minimizing the Gini
        impurity measure.
\end{enumerate}
In our case, we limit the number of trees to 64, the tree depth
to 8, and the maximum number of samples used for training each tree
to 10\%. Tree splitting is based on the Gini impurity. When training
the model, we undersample the negative class,
i.e. we select a random subset of the negative samples
to match the number of positive samples so that the model is not biased toward negative samples.

\subsection{Hyperparameter Tuning}

There are various parameters that must be set before training a random forest model, otherwise known as hyperparameters. Although random forests are generally not overly sensitive to the chosen hyperparameters, the most important are the number of trees and number of features used per tree. Furthermore, although not quite a hyperparameter, a tradeoff between the model complexity and preventing overfitting must be made. This is done by limiting the size of the trees, such as by limiting the tree depth. Lower tree depth also entails faster training.

To optimize the hyperparameters, we temporarily
set aside 10\% of the training sets as validation sets.
We evaluated how the varying parameters affected all of the tasks.
We found that the default recommendation for the number of features per tree was optimal.
Additionally, we experimented with numbers of trees including 32, 64, and 128.
We found that 64 trees provided optimal performance without unnecessary computational overhead. 
Furthermore, we found that a maximum tree depth of 8 provided optimal performance without overfitting. 
Finally, we found that only 10\% of the samples (randomly selected with replacement) per tree
were needed due to the large size of our dataset.
Therefore, we applied this limit to decrease training time.
An added benefit is that there is more disparity between the samples used to train each tree
of the random forests, thus increasing their diversity.

\subsection{Feature Importances}

After building a model using the random forest algorithm,
we use Gini importance to rank the importance of each
feature for that model \citet{RandomForestCART1984}.
In order to calculate the importance of each feature,
we start by calculating the non-normalized
importance $\hat{I}_{f,i}$
of each feature $f$ for each individual tree $i$.
We calculate the sum of the gain $G_j$ (calculated by the random forest algorithm)
weighted by the number
of samples $n_j$
reaching each node $j$ out of the tree's
nodes that split on feature $f$:
\[\hat{I}_{f,i} = \sum_{j} n_j G_j\]
Then, we normalize the feature importances by dividing them by the
sum of all of the importances:
\[I_{f,i} = \frac{\hat{I}_{f,i}}{\sum_i \hat{I}_{f,i}}\]
Finally, we take the mean of the normalized importances
from each tree to get
the overall forest's importance levels for each feature:
\[I_f = \frac{1}{B} \sum_i \hat{I}_{f,i}\]

\subsection{Feature Selection} \label{sec:featureselection}

In order to evaluate the relevance of the features to each task,
we evaluate the model's performance for each task
on a subset of the dataset set aside for testing,
which is not used for training or hyperparameter tuning.
An important consideration is how to split the dataset
into testing and training sets.
If we use random sampling, then that results in the test set being drawn
from the same time period as the training set, which can potentially allow the model to
overfit to the time periods without reducing the performance score on the test set.
On the other hand, the solar wind properties and flux rope occurrence rate
change throughout the solar cycle, so the test set should include samples from throughout
the dataset.
Therefore, we split the data into 20 sequential segments and take the first
20\% of each of those 20 segments for the test set,
and use the remaining to build the training set, which contains 80\% of the
full dataset.

Using the Gini metric of feature importance,
we perform feature selection by iteratively eliminating features in order
of least to most important.
That is, for each task, we remove the least important feature from
the dataset, then re-train and re-evaluate the model with the remaining features.
The performance is evaluated using the Area Under the Receiver-Operator Curve (AUC)
score, which is equivalent to the probability that
a positive sample will receive a higher model output
than a negative sample \citep{hanley1982meaning}.
We continue to do this (without recalculating the feature importances)
until there is only the most important feature remaining.
Then, we plot this result to show
the AUC score for each number of features,
demonstrating how the model performance
decreases when increasingly important features are removed.

A well-known limitation of using Gini importance for feature ranking
is that if some of the features are correlated, then all of the
importance can be given to one of the features at the expense of the other.
A commonly used alternative is permutation importance,
which measures the decrease in the model's performance
if a given feature's values are shuffled randomly.
We have opted to use Gini importance for the following reasons:
(1) Permutation importance is only meaningful if 
the model has high performance, but for some of the cases
in this paper, the random forest classifier was not able to distinguish between
the two categories with good performance. In contrast, Gini importance is still
meaningful for a poorly performing model because it measures the relative information gained
from a particular feature.
(2) Permutation performance can sometimes make the issue of correlations
even worse by reducing the importance of both correlated features.
(3) By experimenting with the use of permutation importance instead of Gini importance,
we have found that for the data used in this study,
the results do not differ significantly
and the conclusions are unaffected.
While more sophisticated methods exist to determine feature importances for correlated features,
the number of features in our study is not that large
and the correlations between them can be understood physically and kept in mind when
interpreting the results.
Therefore, we have opted not to include the results with permutation importance in the paper.
Instead, we discuss the effect of the correlations by physical reasoning
and use them to reach conclusions about the most significant features in Section~\ref{sec:results}.

One issue that may arise when interpreting the feature selection
results is determining how many features one needs
to avoid losing a statistically significant amount of performance.
It is difficult to evaluate this visually when the slope is not steep.
Our solution to this is as follows:
We split the training set into 10 disjoint training folds of equal size
using the same time segmenting procedure as before.
Furthermore, we split the test set into 10 disjoint testing folds of equal size.
Then, we iterate 10 times. For each iteration $i$,
we train the model on the $i$th training fold.
Then, once the model is trained,
we perform a nested iteration 10 times.
For each nested iteration $j$,
we test the model on the $j$th testing fold.
We calculate the AUC score for iteration $ij$ and store it.
At the end, there are $10^2 = 100$ AUC scores.
This is repeated with the top A features and top B features,
yielding two sets of 100 AUC scores for statistical comparison.
We begin with $A = 1$ and $B = 2$, then $A = 2$ and $B = 3$, and so on.
Each time, we perform a Wilcoxon signed-rank test
\citep{wilcoxon1947probability}
to determine whether there is a statistically
significant improvement in performance between
using the top $A$ features and the top $B$ features.
We continue doing this until we encounter a p-value of greater
than 5\%. For example, if this occurs when comparing the top 5 vs the top 6 features
for a particular task,
then the top 5 features are considered significant for that task.

\section{Results} \label{sec:results}

\begin{figure}[htb!]
 \centering \includegraphics[width=.7\textwidth]{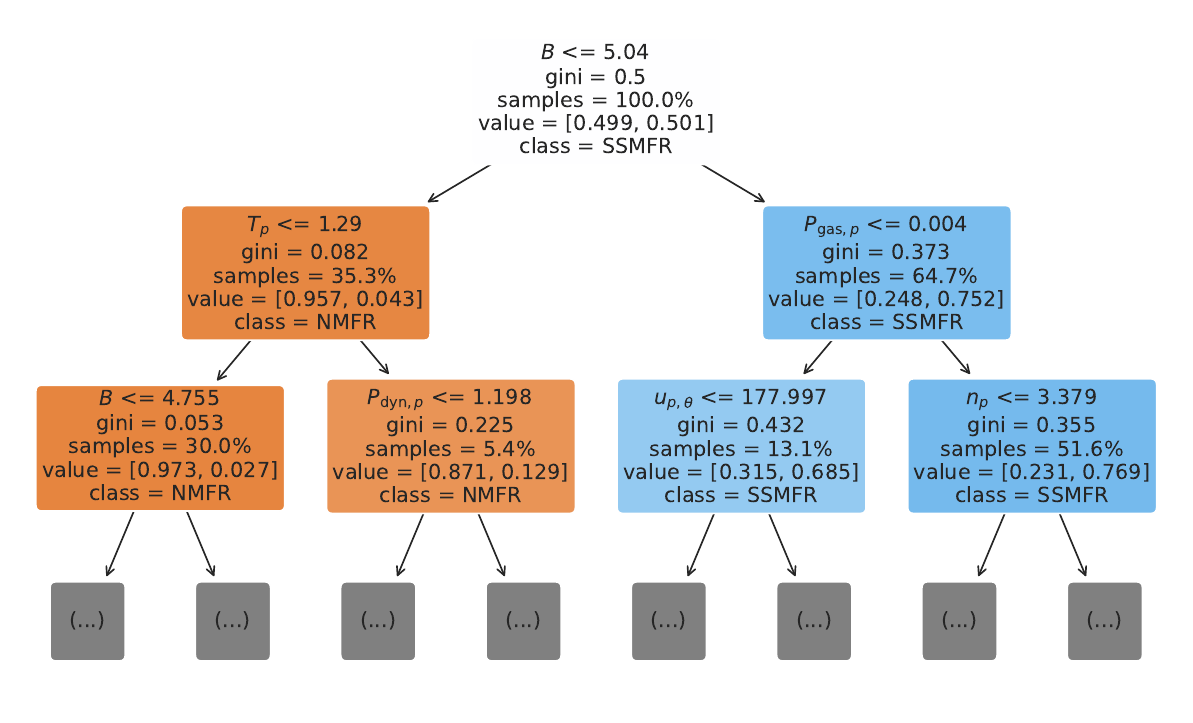}
 \caption{Examples of a decision tree from SSMFR-NMFR. Only the first few levels of the tree are shown. Each task's random forest has many trees; this was selected because it started with the top feature for the respective test. The gini metric shows the impurity after the previous split. Branches to the left result when the condition is satisfied. The ``value'' shows the percentage of NMFRs (left) and percentage of SMFRs (right) assuming that the samples start off balanced.}
\label{fig:trees}
\end{figure}

We trained the classification model
for each of the six tasks using the random forest algorithm
using their respective training sets.
The models themselves consist of many fairly deep trees, so it would be very difficult to visualize them
directly. In Figure~\ref{fig:trees}, we show an example of an individual decision tree
from task SSMFR-NMFR. This illustrates how the decision tree process works
for individual trees. The overall random forest, however,
is based on an ensemble of many trees. Hence,
we use statistical methods below to understand the results.

\begin{deluxetable}{r|cccccc}[htb!]
\label{tab:crosstask}
\tablecaption{Cross-Task Performance}
\tablehead{\colhead{Class} & \colhead{SSMFR-NMFR} & \colhead{LSMFR-NMFR} & \colhead{SSMFR-LSMFR} & \colhead{MC-NMFR} & \colhead{MC-SSMFR} & \colhead{MC-LSMFR}}
\startdata
NMFR & 66.8\% & 66.5\% & - & 88.2\% & - & - \\
SSMFR & 97.0\% & - & 64.0\% & - & 87.6\% & - \\
LSMFR & - & 93.1\% & 63.9\% & - & - & 79.0\% \\
MC & - & - & - & 83.5\% & 82.8\% & 78.5\% \\
\hline
AUC Score & 86.1\% & 85.2\% & 66.2\% & 94.1\% & 92.7\% & 86.4\% \\
\enddata
\end{deluxetable}

We present a detailed
breakdown of the performance of each task across the different
MFR classes in the rest of Table~\ref{tab:crosstask}. The table is generated as follows.
First, we iterate through each class (NMFR, SSMFR, LSMFR, SMFR, MC, MFR).
Then, using the initial models trained on the training sets with all of the features,
predict the class of each sample in the test set belonging to the current class.
We calculate the average predicted class $\langle y' \rangle = \frac{1}{N}\sum y'_i$
where $y'_i$ is $1$ if the model classifies a sample as positive
and $0$ if it classifies a sample as negative.
For tasks where the current class is a subset of the positive class,
we use $\langle y' \rangle$ as the accuracy.
If the current class is a subset of the negative class, we use $1 - \langle y' \rangle$ as the accuracy. If the current class is not a subset of either one, then there is no meaningful accuracy, so
we put dashed lines. The results are recorded in Table~\ref{tab:crosstask}.
Additionally, the resulting AUC scores on the test sets are displayed in the last row of
Table~\ref{tab:crosstask}
With the exception of SSMFR-LSMFR,
all of them had around 90\% AUC, with MC-NMFR exceeding 94\%.
The MC-related tasks performed better than the SMFR-related tasks.
Higher AUC scores indicate that the model is able to assign
a higher prediction value to a true sample over a negative sample more often,
and thus the features given to it have more predictive power.
The baseline AUC score for random guessing is 0.5, so in all cases,
the model had significant (if not good) performance.
The fact that the model can give meaningful probability estimates indicates that
there is a significant correspondence between the features that it found and
the labels of each task.

The poor performance of SSMFR-LSMFR is important.
If big SMFRs tended to be more similar to MCs, there should have been a big physical difference between LSMFR and SSMFR and SSMFR-LSMFR should have scored well. In fact, MC-LSMFR does perform worse than MC-SSMFR, but SSMFR-LSMFR performs worse than either one of them, suggesting big SMFRs have more in common with small SMFRs than MCs.
The physical properties exhibit only slight differences, as we will see below.
We will also explore the reasons that MC-NMFR, SSMFR-NMFR, and LSMFR-NMFR perform well.
\begin{deluxetable*}{r|cccccc}[htb!]
\label{tab:ranking}
\tablecaption{Feature Importances (Significant Features in Red)}
\tablehead{\colhead{Rank} & \colhead{SSMFR-NMFR} & \colhead{LSMFR-NMFR} & \colhead{SSMFR-LSMFR} & \colhead{MC-NMFR} & \colhead{MC-SSMFR} & \colhead{MC-LSMFR}}
\startdata
1 & \color{red}{$B$} & \color{red}{$B$} & \color{red}{$\beta_p$} & \color{red}{$\beta_p$} & \color{red}{$\beta_p$} & \color{red}{$\beta_p$}\\
2 & \color{red}{$M_A$} & \color{red}{$M_A$} & \color{red}{$T_p/T_{ex}$} & \color{red}{$B$} & \color{red}{$T_p/T_{ex}$} & \color{red}{$T_p/T_{ex}$}\\
3 & $v_A$ & \color{red}{$v_A$} & \color{red}{$P_{\mathrm{gas},p}$} & \color{red}{$M_A$} & \color{red}{$P_{\mathrm{gas},p}$} & \color{red}{$M_A$}\\
4 & $P_{\mathrm{gas},p}$ & \color{red}{$P_{\mathrm{dyn},p}$} & \color{red}{$u_p$} & \color{red}{$v_A$} & \color{red}{$M_A$} & \color{red}{$B$}\\
5 & $P_{\mathrm{dyn},p}$ & \color{red}{$n_p$} & \color{red}{$B$} & \color{red}{$T_p/T_{ex}$} & \color{red}{$B$} & \color{red}{$P_{\mathrm{gas},p}$}\\
6 & $n_p$ & \color{red}{$\beta_p$} & \color{red}{$n_{\alpha}/n_p$} & $P_{\mathrm{gas},p}$ & \color{red}{$v_A$} & $v_A$\\
7 & $T_p/T_{ex}$ & $P_{\mathrm{gas},p}$ & \color{red}{$v_A$} & $n_p$ & $P_{\mathrm{dyn},p}$ & $n_p$\\
8 & $\beta_p$ & $T_p/T_{ex}$ & \color{red}{$n_p$} & $P_{\mathrm{dyn},p}$ & $n_p$ & $P_{\mathrm{dyn},p}$\\
9 & $u_p$ & $u_p$ & $M_A$ & $u_p$ & $n_{\alpha}/n_p$ & $n_{\alpha}/n_p$\\
10 & $B_{\theta}$ & $u_{p,\phi}$ & $P_{\mathrm{dyn},p}$ & $n_{\alpha}/n_p$ & $u_p$ & $u_p$\\
11 & $u_{p,\theta}$ & $B_{\theta}$ & $B_{\phi}$ & $B_{\theta}$ & $u_{p,\phi}$ & $B_{\phi}$\\
12 & $u_{p,\phi}$ & $n_{\alpha}/n_p$ & $u_{p,\theta}$ & $B_{\phi}$ & $u_{p,\theta}$ & $u_{p,\theta}$\\
13 & $n_{\alpha}/n_p$ & $B_{\phi}$ & $u_{p,\phi}$ & $u_{p,\phi}$ & $B_{\phi}$ & $u_{p,\phi}$\\
14 & $B_{\phi}$ & $u_{p,\theta}$ & $B_{\theta}$ & $u_{p,\theta}$ & $B_{\theta}$ & $B_{\theta}$\\
\enddata
\end{deluxetable*}
\begin{figure*}[htb!]
 \centering
 \includegraphics[width=0.49\textwidth]{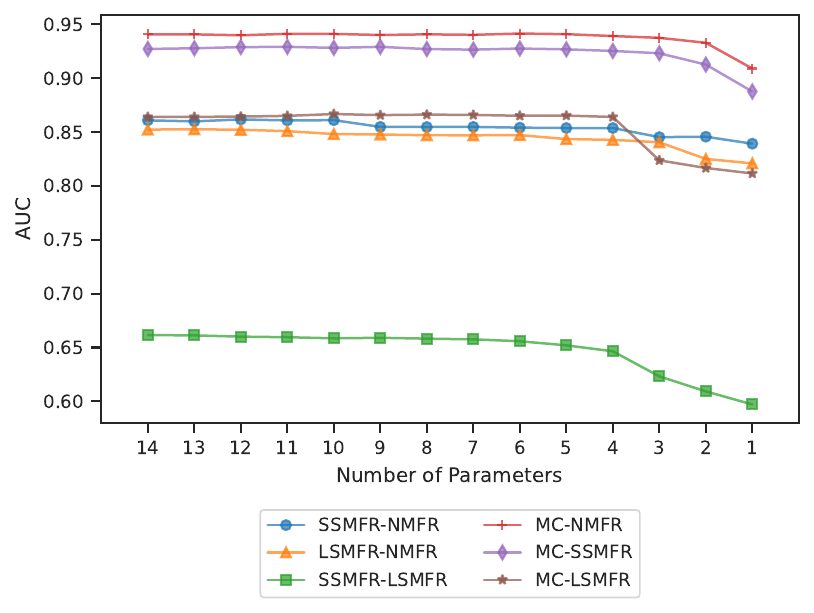}
 \includegraphics[width=0.49\textwidth]{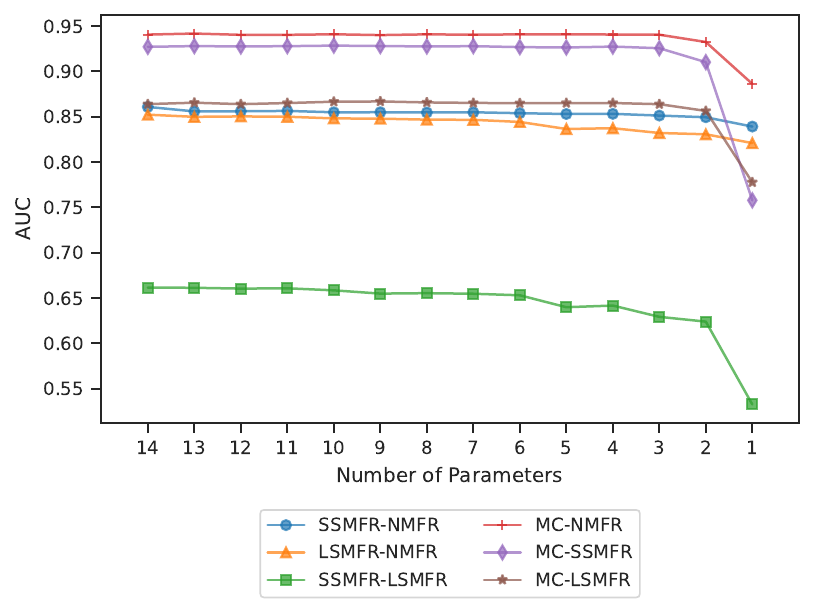}
\caption{(a) Feature elimination results generated using each individual task's
            feature ranking. (b) Same as (a), except rather than dropping features
            based on each task's ranking, drop in order of average importance across tasks.}
\label{fig:selection}
\end{figure*}

After training each task's model, we calculate the feature importances for each task,
then use it to rank the features, recorded in Table~\ref{tab:ranking}.
Finally, we generated the feature selection curves for each task.
The least important feature is dropped iteratively until only the most important feature
remains, and with each subset of features, the model is re-trained and the AUC score is calculated.
Figure~\ref{fig:selection} (a) shows the plot of the results for each task.
In order to determine which features were significant for each task,
we performed the Wilcoxon tests in the manner described in Section~\ref{sec:featureselection}.
Using this information, we colored the significant features red in Table~\ref{tab:ranking}.
We also sorted the features by their average importance across tasks
and then conducted the same process except using this new cross-task ordering.
The resulting feature elimination plot is Figure~\ref{fig:selection} (b).
The top 5 features across tasks were $B$, $\beta_p$, $T_p/T_\mathrm{exp}$, $M_A$, and $P_\mathrm{gas}$.

\begin{figure}[htb!]
 \centering \includegraphics[width=.9\textwidth]{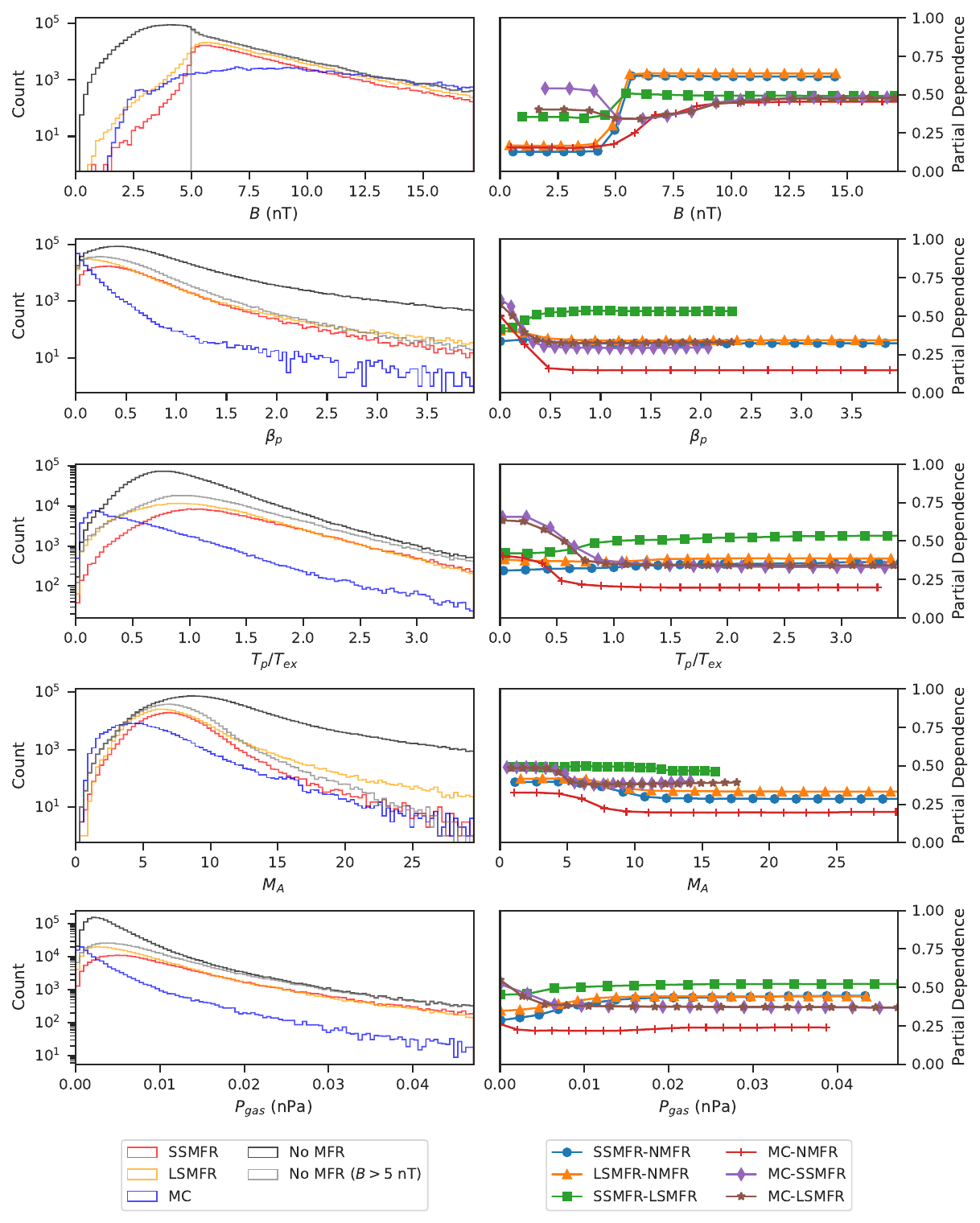}
 \caption{Histogram and partial dependency plots for the top 5 most important features across tasks.}
\label{fig:histograms_and_pdps}
\end{figure}

The feature ranking tells us which features
are significant for distinguishing a sample's class,
but not what values correspond to a higher probability.
These can be understood with histograms and partial dependency plots (PDPs), plotted in Figure~\ref{fig:histograms_and_pdps}.
To generate the PDPs,
each model is trained with all features using the task's training set,
then 10,000 random samples from the task's test set are selected.
Take magnetic field strength $B$ as an example.
For each point on the PDP, the magnetic field strength $B$ of each selected sample is changed to a new value $B'$,
and then all the samples are passed through the task's model.
The mean resulting probability prediction is represented by a point on the PDP.
The change of the value of the PDP shows how
the model's output depends on a given parameter.

For both SSMFR-NMFR and LSMFR-NMFR, the top feature is the magnetic field strength $B$.
This implies that the magnetic field strength stands out the most during the SMFRs in the event list
compared to other features. The comparison of their distributions in Figure~\ref{fig:histograms_and_pdps}
reveals a sharp change centered at \SI{5}{\nano\tesla}:
Only 2.7\% of samples with $B < \SI{5}{\nano\tesla}$ contain an SMFR, whereas
47.6\% of samples with $B \geq \SI{5}{\nano\tesla}$ contain an SMFR.
Apparently, this is due to the fact that \citet{hu2018automated}
excluded any MFR event candidates with an average magnetic field strength below
a \SI{5}{\nano\tesla} threshold in order to exclude random fluctuations.
We have also included the distribution for the non-MFR samples with the same
threshold applied, which results in a very similar distribution to the SMFR distribution.
The PDP shows a very sharp increase after \SI{5}{\nano\tesla}
from approximately 0.1 or less to a nearly constant value for SMFR tasks.
In contrast to the PDP for the previously discussed tasks,
the MC tasks
have a smooth increase of probability with magnetic field strength.
Interestingly, MC-SSMFR and MC-LSMFR have an \emph{increase}
in probability below 5 nT,
suggesting that at values below the threshold, 
the model gives a higher chance of the sample being an MC than an SMFR
even though MCs tend to have higher magnetic field strength than SMFRs.
Overall, it seems that there may be many SMFRs with lower magnetic field strength
and that the threshold has imposed a selection bias.

For the MC-related tasks,
the most important feature is proton beta.
In fact, judging by the PDP,
$\beta_p$ is the main factor in distinguishing MCs from other categories.
It is well-known that the proton beta tends to be very low for ICMEs
(e.g. \citet{liu2005statistical}).
MCs tend to have near-0 values
which only occur rarely in the background solar wind.
However, the tail of the distribution cannot be ignored,
since observational studies show that not all MCs have low proton beta
for their entire duration (e.g. \citet{pal2022eruption}).

SSMFRs and LSMFRs also tend to have lower proton beta than NMFRs,
although the difference is not nearly as pronounced as in the case of MCs.
The distribution of SMFR proton beta has been analyzed by \citet{hu2018automated}.
The smaller difference between MC and SMFR proton beta compared to MC and non-MFR
may explain the lower AUC score for MC-SSMFR and MC-LSMFR.
Likewise, it appears that LSMFRs have slightly lower $\beta_p$ than SSMFRs, so MC-LSMFR performs worse than MC-SSMFR. However, the shape of the distribution for LSMFRs, SSMFRs, and NMFRs is virtually the same, whereas MC-LSMFR has a completely different distribution shape, so this does not necessarily indicate a similarity between the plasma properties of LSMFRs and MCs.
In fact, both SSMFRs and LSMFRs have almost the same
$\beta_p$ distribution as NMFRs that have $B > \SI{5}{\nano\tesla}$, suggesting they both share the distribution of the background solar wind.

Most features besides $B$ do not appear to be significantly different for SMFRs
based on the histograms compared to
NMFRs (with the 5 nT threshold applied).
$M_A$ was given high importance for SMFR-related tasks,
but does not appear to affect the PDP much compared to $B$.
Since it is highly correlated with $B$ Figure~\ref{fig:correlation}, that is probably what resulted in its high ranking. In contrast,
MCs have significantly different shape of the 
 $M_A$ distribution, but since the overall distribution
 has too much overlap in terms of the range of values, it was not ranked highly in terms of Gini importance.

Beta, temperature, and gas pressure appear to play an important role for distinguishing MCs from NMFRs and LSMFRs from SSMFRs.
As \citet{hu2018automated} pointed out,
smaller SMFRs have a wide range of temperatures whereas larger SMFRs have lower temperatures.
That finding is consistent with the PDP, which shows that SSMFRs are more likely to have higher temperatures than LSMFRs.
But even though LSMFRs have low temperature,
their temperature distribution is shaped completely different from MCs. Unlike MCs, the distribution of LSMFR temperature is similar to both SSMFR and NSMFR temperature. Accordingly, temperature only plays a very strong role for MC-related tasks in the PDP.
Gas pressure, related to temperature by $P = nk_BT$, tends to be very low for MCs compared to SMFRs, LSMFRs, or NMFRs, which all have a similar distribution of gas pressure. However, larger SMFRs appear to have slightly lower gas pressure.

\section{Discussion and Conclusions} \label{sec:discussion}

In this paper, we ranked the informativeness of various physical properties of the solar wind properties
with regards to various MFR-related categorizations (SSMFR, LSMFR, MC, and NMFR) using a machine learning feature selection technique.
First, we generated a dataset using measurements from the \emph{Wind} spacecraft.
Then, we used 6 binary classification to compare the point-in-time physical properties of each unique pair of
categories.
For each task, we used the random forest algorithm to build an ensemble of decision trees to distinguish
the positive from negative samples and then used the normalized average decrease in the impurity of each feature
to rank the features.
After that, we applied feature selection by plotting the change in model performance with
increasingly reduced sets of top features used for training and classification.

Our results demonstrated that MCs tend to have very distinctive solar wind properties compared to the absence of any MFR,
such that an excellent probability estimate of a given timestamp corresponding to an MC instead of a non-MFR
can be given based on plasma and magnetic field measurements from a single point in time. 
The proton beta alone is able to give an AUC score of over 0.9 for distinguishing MCs from non-MFRs whereas the top 3 features
together ($\beta_p$, $B$, and $v_A$) result in an AUC score of close to 0.94. This makes sense
because it is known that intervals of both low $\beta_p$ and $v_A$
are most common in the solar wind during ICMEs \citep{gosling2013magnetic},
and that MCs tend to have elevated magnetic field strength $B$ \citep{nieves2018icmecatalog}.

We found that SMFRs in the catalog can be distinguished from non-MFRs
with significant performance.
However, the primary distinguishing feature is just the $B > \SI{5}{\nano\tesla}$ threshold
used by \citet{hu2018automated} to make the catalog.
As a result, the plasma properties that differ
significantly in distribution from the non-MFRs---$v_A$,
$M_A$, and $P_\mathrm{gas}$ in particular---are
strongly
correlated with $B$.
Therefore, it seems that SMFRs share the same distribution as the background solar wind
in terms of instantaneous physical properties.
A potential area for future research is to detect flux ropes from multiple-spacecraft observations,
eliminating the need for the threshold, since one can be more confident of a detected flux rope not being a mere fluctuation.

Our comparison between SMFRs shorter or longer than an hour in duration (corresponding to small and large: SMFRs tend to be Parker spiral aligned \citep{hu2018automated}, so duration is a good proxy for size) suggests that there is almost no different between their physical properties. Even though, as noted by \citet{hu2018automated},
larger SMFRs have lower proton temperature, proton beta, and gas pressure,
the difference is small compared to how low the temperature and beta of MCs is.
Larger SMFRs have more in common with smaller SMFRs and the background solar wind than with MCs.
The main difference in physical properties between larger SMFRs and smaller SMFRs is that larger SMFRs
have slightly lower gas pressure, temperature, and proton beta. However, the distributions of these parameters have the same shape as the distributions of smaller events, whereas both are very different from MCs.
Additionally, the magnetic field fluctuates more in smaller flux ropes than larger ones, although that is due to their structure's size, not their plasma properties.

Based on our results, we hypothesize that most SMFRs at 1 au
share the same plasma properties as the ambient solar wind.
This is in contrast to the typical description of SMFRs as transient phenomena
like MCs, which are distinct plasma regimes in the solar wind.
Considering that the SMFR events in the current catalog with the \SI{5}{\nano\tesla}
threshold applied already span 26\% of the full time, i.e. SMFRs are present at 1 au
over one quarter of the time, this makes sense.
In the context of the flux-tube picture of the solar wind by \citet{borovsky2008flux},
the above discussion suggests that the flux tubes filling the solar wind are often twisted,
which may have important ramifications for the interaction between the solar wind and the magnetosphere.
It also provides further support for most SMFRs at 1 au having a local origin
within the solar wind, e.g. by turbulence \citep{hu2018automated}.
However, this study is based on the bulk fluid properties of the solar wind. Analysis of the microscopic
properties such as particle
distribution functions may
yield different conclusions
regarding the origin of the SMFRs.

Future works may extend this feature ranking methodology to data from other spacecraft
in different parts of the heliosphere and compare the change in results based on variations in ecliptic latitude
and distance from the Sun. Furthermore, this methodology can be applied to other phenomena
with various in-situ time series measurements of which the relative importances are of interest. 

\section*{Acknowledgments}
We thank NASA's Space Physics Data Facility
for providing the \emph{Wind} spacecraft data,
and we thank
the teams at the University of Alabama in Huntsville (UAH) and NASA
for the SMFR (available at \verb|https://fluxrope.info|)
and MC (available at \verb|https://wind.nasa.gov/ICME_catalog/ICME_catalog_viewer.php|)
catalogs, respectively.
We acknowledge the support of NASA grant 80NSSC20K1282
and NSF grants AGS-1927578, AGS-2229064, and OPP-2032421.
The figures and results in this study were generated using Python
and Python libraries including
matplotlib \citep{matplotlib}, SciKit-Learn \citep{scikit-learn}, numpy \citep{numpy},
scipy \citep{scipy},
and pandas \citep{pandas}.

\bibliography{refs}{}

\begin{thebibliography}{}
\expandafter\ifx\csname natexlab\endcsname\relax\def\natexlab#1{#1}\fi
\providecommand{\url}[1]{\href{#1}{#1}}
\providecommand{\dodoi}[1]{doi:~\href{http://doi.org/#1}{\nolinkurl{#1}}}
\providecommand{\doeprint}[1]{\href{http://ascl.net/#1}{\nolinkurl{http://ascl.net/#1}}}
\providecommand{\doarXiv}[1]{\href{https://arxiv.org/abs/#1}{\nolinkurl{https://arxiv.org/abs/#1}}}

\bibitem[{Abduallah {et~al.}(2022)Abduallah, Jordanova, Liu, Li, Wang, \& Wang}]{abduallah2022predicting}
Abduallah, Y., Jordanova, V.~K., Liu, H., {et~al.} 2022, The Astrophysical Journal Supplement Series, 260, 16.
\newblock \url{https://doi.org/10.3847/1538-4365/ac5f56}

\bibitem[{Alpaydin(2020)}]{alpaydin2020introduction}
Alpaydin, E. 2020, Introduction to machine learning (MIT press)

\bibitem[{Baker {et~al.}(2004)Baker, Daly, Daglis, Kappenman, \& Panasyuk}]{baker2004effects}
Baker, D.~N., Daly, E., Daglis, I., Kappenman, J.~G., \& Panasyuk, M. 2004, Space Weather, 2.
\newblock \url{https://doi.org/10.1029/2003SW000044}

\bibitem[{Borovsky(2008)}]{borovsky2008flux}
Borovsky, J.~E. 2008, Journal of Geophysical Research: Space Physics, 113.
\newblock \url{https://doi.org/10.1029/2007JA012684}

\bibitem[{{Breiman} {et~al.}(1984){Breiman}, {Friedman}, \& {Olshen}}]{RandomForestCART1984}
{Breiman}, L., {Friedman}, J.~H., \& {Olshen}, R.~A. 1984, Classification and Regression Trees (Boca Raton, Florida: Chapman and Hall/CRC).
\newblock \url{https://doi.org/10.1201/9781315139470}

\bibitem[{Burlaga {et~al.}(1981)Burlaga, Sittler, Mariani, \& Schwenn}]{burlaga1981magnetic}
Burlaga, L., Sittler, E., Mariani, F., \& Schwenn, a.~R. 1981, Journal of Geophysical Research: Space Physics, 86, 6673.
\newblock \url{https://doi.org/10.1029/JA086iA08p06673}

\bibitem[{Camporeale {et~al.}(2017)Camporeale, Car{\`e}, \& Borovsky}]{camporeale2017classification}
Camporeale, E., Car{\`e}, A., \& Borovsky, J.~E. 2017, Journal of Geophysical Research: Space Physics, 122, 10, \dodoi{https://doi.org/10.1002/2017JA024383}

\bibitem[{Chen(2017)}]{chen2017physics}
Chen, J. 2017, Physics of Plasmas, 24, 090501.
\newblock \url{https://doi.org/10.1063/1.4993929}

\bibitem[{Chen \& Hu(2020)}]{chen2020effects}
Chen, Y., \& Hu, Q. 2020, The Astrophysical Journal, 894, 25.
\newblock \url{https://doi.org/10.3847/1538-4357/ab8294}

\bibitem[{Chen \& Hu(2022)}]{chen2022small}
---. 2022, The Astrophysical Journal, 924, 43.
\newblock \url{https://doi.org/10.3847/1538-4357/ac3487}

\bibitem[{Chen {et~al.}(2019)Chen, Hu, \& le~Roux}]{chen2019analysis}
Chen, Y., Hu, Q., \& le~Roux, J.~A. 2019, The Astrophysical Journal, 881, 58.
\newblock \url{https://doi.org/10.3847/1538-4357/ab2ccf}

\bibitem[{Chen {et~al.}(2020)Chen, Hu, Zhao, Kasper, Bale, Korreck, Case, Stevens, Bonnell, Goetz, Harvey, Klein, Larson, Livi, MacDowall, Malaspina, Pulupa, \& Whittlesey}]{yuchen2020_gs_psp_first_two}
Chen, Y., Hu, Q., Zhao, L., {et~al.} 2020, The Astrophysical Journal, 903, 76.
\newblock \url{https://doi.org/10.3847/1538-4357/abb820}

\bibitem[{dos Santos {et~al.}(2020)dos Santos, Narock, Nieves-Chinchilla, Nu{\~n}ez, \& Kirk}]{dos2020identifying}
dos Santos, L.~F., Narock, A., Nieves-Chinchilla, T., Nu{\~n}ez, M., \& Kirk, M. 2020, Solar Physics, 295, 1.
\newblock \url{https://doi.org/10.1007/s11207-020-01697-x}

\bibitem[{Gibson(2018)}]{gibson2018prominencereview}
Gibson, S.~E. 2018, Living reviews in solar physics, 15, 1

\bibitem[{Gosling \& Phan(2013)}]{gosling2013magnetic}
Gosling, J.~T., \& Phan, T.~D. 2013, The Astrophysical Journal, 763, L39.
\newblock \url{https://doi.org/10.1088/2041-8205/763/2/l39}

\bibitem[{Hanley \& McNeil(1982)}]{hanley1982meaning}
Hanley, J.~A., \& McNeil, B.~J. 1982, Radiology, 143, 29.
\newblock \url{https://doi.org/10.1148/radiology.143.1.7063747}

\bibitem[{Harris {et~al.}(2020)Harris, Millman, van~der Walt, Gommers, Virtanen, Cournapeau, Wieser, Taylor, Berg, Smith, Kern, Picus, Hoyer, van Kerkwijk, Brett, Haldane, del R{\'{i}}o, Wiebe, Peterson, G{\'{e}}rard-Marchant, Sheppard, Reddy, Weckesser, Abbasi, Gohlke, \& Oliphant}]{numpy}
Harris, C.~R., Millman, K.~J., van~der Walt, S.~J., {et~al.} 2020, Nature, 585, 357.
\newblock \url{https://doi.org/10.1038/s41586-020-2649-2}

\bibitem[{Howard \& Tappin(2009)}]{howard2009interplanetary}
Howard, T.~A., \& Tappin, S.~J. 2009, Space science reviews, 147, 31.
\newblock \url{https://doi.org/10.1007/s11214-009-9542-5}

\bibitem[{Hu {et~al.}(2019)Hu, Chen, \& le~Roux}]{hu2019radial}
Hu, Q., Chen, Y., \& le~Roux, J. 2019, Journal of Physics: Conference Series, 1332, 012005.
\newblock \url{https://doi.org/10.1088/1742-6596/1332/1/012005}

\bibitem[{Hu {et~al.}(2014)Hu, Qiu, Dasgupta, Khare, \& Webb}]{hu2014structures}
Hu, Q., Qiu, J., Dasgupta, B., Khare, A., \& Webb, G.~M. 2014, The Astrophysical Journal, 793, 53.
\newblock \url{https://doi.org/10.1088/0004-637x/793/1/53}

\bibitem[{Hu {et~al.}(2018)Hu, Zheng, Chen, le~Roux, \& Zhao}]{hu2018automated}
Hu, Q., Zheng, J., Chen, Y., le~Roux, J., \& Zhao, L. 2018, The Astrophysical Journal Supplement Series, 239, 12.
\newblock \url{https://doi.org/10.3847/1538-4365/aae57d}

\bibitem[{Hunter(2007)}]{matplotlib}
Hunter, J.~D. 2007, Computing in Science \& Engineering, 9, 90.
\newblock \url{https://doi.org/10.1109/MCSE.2007.55}

\bibitem[{Kilpua {et~al.}(2017)Kilpua, Koskinen, \& Pulkkinen}]{kilpua2017coronal}
Kilpua, E., Koskinen, H.~E., \& Pulkkinen, T.~I. 2017, Living Reviews in Solar Physics, 14, 1.
\newblock \url{https://doi.org/10.1007/s41116-017-0009-6}

\bibitem[{Klein \& Burlaga(1982)}]{klein1982interplanetary}
Klein, L., \& Burlaga, L. 1982, Journal of Geophysical Research: Space Physics, 87, 613, \dodoi{https://doi.org/10.1029/JA087iA02p00613}

\bibitem[{Lavraud {et~al.}(2013)Lavraud, Larroque, Budnik, G{\'e}not, Borovsky, Dunlop, Foullon, Hasegawa, Jacquey, Nykyri, {et~al.}}]{lavraud2013asymmetry}
Lavraud, B., Larroque, E., Budnik, E., {et~al.} 2013, Journal of Geophysical Research: Space Physics, 118, 1089.
\newblock \url{https://doi.org/10.1002/jgra.50145}

\bibitem[{Lepping {et~al.}(1995)Lepping, Ac{\~u}na, Burlaga, Farrell, Slavin, Schatten, Mariani, Ness, Neubauer, Whang, {et~al.}}]{lepping1995wind}
Lepping, R., Ac{\~u}na, M., Burlaga, L., {et~al.} 1995, Space Science Reviews, 71, 207, \dodoi{https://doi.org/10.1007/BF00751330}

\bibitem[{Li {et~al.}(2020)Li, Wang, Tu, \& Xu}]{li2020machine}
Li, H., Wang, C., Tu, C., \& Xu, F. 2020, Earth and Space Science, 7, e2019EA000997, \dodoi{https://doi.org/10.1029/2019EA000997}

\bibitem[{Liu {et~al.}(2017)Liu, Deng, Wang, \& Wang}]{liu2017rfsolarflare}
Liu, C., Deng, N., Wang, J. T.~L., \& Wang, H. 2017, The Astrophysical Journal, 843, 104.
\newblock \url{https://doi.org/10.3847/1538-4357/aa789b}

\bibitem[{Liu {et~al.}(2020)Liu, Liu, Wang, \& Wang}]{liu2020predicting}
Liu, H., Liu, C., Wang, J. T.~L., \& Wang, H. 2020, The Astrophysical Journal, 890, 12.
\newblock \url{https://doi.org/10.3847/1538-4357/ab6850}

\bibitem[{Liu(2020)}]{liu2020fluxropecorona}
Liu, R. 2020, Research in Astronomy and Astrophysics, 20, 165.
\newblock \url{https://doi.org/10.1088/1674-4527/20/10/165}

\bibitem[{Liu {et~al.}(2005)Liu, Richardson, \& Belcher}]{liu2005statistical}
Liu, Y., Richardson, J., \& Belcher, J. 2005, Planetary and Space Science, 53, 3.
\newblock \url{https://doi.org/10.1016/j.pss.2004.09.023}

\bibitem[{Moldwin {et~al.}(2000)Moldwin, Ford, Lepping, Slavin, \& Szabo}]{moldwin2000smfr}
Moldwin, M., Ford, S., Lepping, R., Slavin, J., \& Szabo, A. 2000, Geophysical research letters, 27, 57.
\newblock \url{https://doi.org/10.1029/1999GL010724}

\bibitem[{Moldwin {et~al.}(1995)Moldwin, Phillips, Gosling, Scime, McComas, Bame, Balogh, \& Forsyth}]{moldwin1995ulysses}
Moldwin, M., Phillips, J., Gosling, J., {et~al.} 1995, Journal of Geophysical Research: Space Physics, 100, 19903.
\newblock \url{https://doi.org/10.1029/95JA01123}

\bibitem[{Mullan \& Smith(2006)}]{mullan2006solar}
Mullan, D., \& Smith, C. 2006, Solar Physics, 234, 325.
\newblock \url{https://doi.org/10.1007/s11207-006-2077-y}

\bibitem[{Narock {et~al.}(2022)Narock, Narock, Dos~Santos, \& Nieves-Chinchilla}]{narock2022identification}
Narock, T., Narock, A., Dos~Santos, L.~F., \& Nieves-Chinchilla, T. 2022, Frontiers in Astronomy and Space Sciences, 9, 27.
\newblock \url{https://doi.org/10.3389/fspas.2022.838442}

\bibitem[{Nguyen {et~al.}(2019)Nguyen, Aunai, Fontaine, Pennec, Bossche, Jeandet, Bakkali, Vignoli, \& Blancard}]{nguyen2019automatic}
Nguyen, G., Aunai, N., Fontaine, D., {et~al.} 2019, The Astrophysical Journal, 874, 145, \dodoi{https://doi.org/10.3847/1538-4357/ab0d24}

\bibitem[{Nieves-Chinchilla {et~al.}(2018)Nieves-Chinchilla, Vourlidas, Raymond, Linton, Al-Haddad, Savani, Szabo, \& Hidalgo}]{nieves2018icmecatalog}
Nieves-Chinchilla, T., Vourlidas, A., Raymond, J., {et~al.} 2018, Solar Physics, 293, 1.
\newblock \url{https://doi.org/10.1007/s11207-018-1247-z}

\bibitem[{Ogilvie {et~al.}(1995)Ogilvie, Chornay, Fritzenreiter, Hunsaker, Keller, Lobell, Miller, Scudder, Sittler, Torbert, {et~al.}}]{ogilvie1995swe}
Ogilvie, K., Chornay, D., Fritzenreiter, R., {et~al.} 1995, Space Science Reviews, 71, 55, \dodoi{https://doi.org/10.1007/BF00751326}

\bibitem[{Pal {et~al.}(2022)Pal, Lynch, Good, Palmerio, Asvestari, Pomoell, Stevens, \& Kilpua}]{pal2022eruption}
Pal, S., Lynch, B.~J., Good, S.~W., {et~al.} 2022, Frontiers in Astronomy and Space Sciences, 136.
\newblock \url{https://doi.org/10.3389/fspas.2022.903676}

\bibitem[{Pedregosa {et~al.}(2011)Pedregosa, Varoquaux, Gramfort, Michel, Thirion, Grisel, Blondel, Prettenhofer, Weiss, Dubourg, Vanderplas, Passos, Cournapeau, Brucher, Perrot, \& Duchesnay}]{scikit-learn}
Pedregosa, F., Varoquaux, G., Gramfort, A., {et~al.} 2011, Journal of Machine Learning Research, 12, 2825

\bibitem[{Raheem {et~al.}(2021)Raheem, Cavus, Coban, Kinaci, Wang, \& Wang}]{raheem2021investigation}
Raheem, A.-u., Cavus, H., Coban, G.~C., {et~al.} 2021, Monthly Notices of the Royal Astronomical Society, 506, 1916.
\newblock \url{https://doi.org/10.1093/mnras/stab1816}

\bibitem[{Reiss {et~al.}(2021)Reiss, Möstl, Bailey, Rüdisser, Amerstorfer, Amerstorfer, Weiss, Hinterreiter, \& Windisch}]{reiss2021bz}
Reiss, M.~A., Möstl, C., Bailey, R.~L., {et~al.} 2021, Space Weather, 19.
\newblock \url{https://doi.org/10.1029/2021SW002859}

\bibitem[{Richardson \& Cohen(2021)}]{richardson2021seasonal}
Richardson, D.~K., \& Cohen, M.~B. 2021, Journal of Geophysical Research: Space Physics, 126.
\newblock \url{https://doi.org/10.1029/2021JA029689}

\bibitem[{Richardson \& Cane(1995)}]{richardson1995regions}
Richardson, I., \& Cane, H. 1995, Journal of Geophysical Research: Space Physics, 100, 23397.
\newblock \url{https://doi.org/10.1029/95JA02684}

\bibitem[{Roberts {et~al.}(2020)Roberts, Karimabadi, Sipes, Ko, \& Lepri}]{roberts2020objectively}
Roberts, D.~A., Karimabadi, H., Sipes, T., Ko, Y.-K., \& Lepri, S. 2020, The Astrophysical Journal, 889, 153.
\newblock \url{https://doi.org/10.3847/1538-4357/ab5a7a}

\bibitem[{Rouillard {et~al.}(2010)Rouillard, Davies, Lavraud, Forsyth, Savani, Bewsher, Brown, Sheeley, Davis, Harrison, {et~al.}}]{rouillard2010intermittent}
Rouillard, A., Davies, J., Lavraud, B., {et~al.} 2010, Journal of Geophysical Research: Space Physics, 115, \dodoi{https://doi.org/10.1029/2009JA014471}

\bibitem[{Rouillard {et~al.}(2011)Rouillard, Sheeley, Cooper, Davies, Lavraud, Kilpua, Skoug, Steinberg, Szabo, Opitz, {et~al.}}]{rouillard2011solar}
Rouillard, A., Sheeley, N., Cooper, T., {et~al.} 2011, The Astrophysical Journal, 734, 7, \dodoi{https://doi.org/10.1088/0004-637X/734/1/7}

\bibitem[{Sanchez-Diaz {et~al.}(2019)Sanchez-Diaz, Rouillard, Lavraud, Kilpua, \& Davies}]{sanchez2019situ}
Sanchez-Diaz, E., Rouillard, A., Lavraud, B., Kilpua, E., \& Davies, J. 2019, The Astrophysical Journal, 882, 51, \dodoi{https://doi.org/10.3847/1538-4357/ab341c}

\bibitem[{Sanchez-Diaz {et~al.}(2017{\natexlab{a}})Sanchez-Diaz, Rouillard, Davies, Lavraud, Pinto, \& Kilpua}]{sanchez2017temporal}
Sanchez-Diaz, E., Rouillard, A.~P., Davies, J.~A., {et~al.} 2017{\natexlab{a}}, The Astrophysical Journal, 851, 32, \dodoi{https://doi.org/10.3847/1538-4357/aa98e2}

\bibitem[{Sanchez-Diaz {et~al.}(2017{\natexlab{b}})Sanchez-Diaz, Rouillard, Davies, Lavraud, Sheeley, Pinto, Kilpua, Plotnikov, \& Genot}]{sanchez2017observational}
---. 2017{\natexlab{b}}, The Astrophysical Journal Letters, 835, L7, \dodoi{https://doi.org/10.3847/2041-8213/835/1/L7}

\bibitem[{Sundberg {et~al.}(2017)Sundberg, Burgess, Scholer, Masters, \& Sulaiman}]{sundberg2017dynamics}
Sundberg, T., Burgess, D., Scholer, M., Masters, A., \& Sulaiman, A.~H. 2017, The Astrophysical Journal, 836, L4.
\newblock \url{https://doi.org/10.3847/2041-8213/836/1/l4}

\bibitem[{Vech \& Malaspina(2021)}]{vech2021novel}
Vech, D., \& Malaspina, D.~M. 2021, Journal of Geophysical Research: Space Physics, 126.
\newblock \url{https://doi.org/10.1029/2021JA029567}

\bibitem[{Virtanen {et~al.}(2020)Virtanen, Gommers, Oliphant, Haberland, Reddy, Cournapeau, Burovski, Peterson, Weckesser, Bright, {van der Walt}, Brett, Wilson, Millman, Mayorov, Nelson, Jones, Kern, Larson, Carey, Polat, Feng, Moore, {VanderPlas}, Laxalde, Perktold, Cimrman, Henriksen, Quintero, Harris, Archibald, Ribeiro, Pedregosa, {van Mulbregt}, \& {SciPy 1.0 Contributors}}]{scipy}
Virtanen, P., Gommers, R., Oliphant, T.~E., {et~al.} 2020, Nature Methods, 17, 261

\bibitem[{{W}es {M}c{K}inney(2010)}]{pandas}
{W}es {M}c{K}inney. 2010, in {P}roceedings of the 9th {P}ython in {S}cience {C}onference, ed. {S}t\'efan van~der {W}alt \& {J}arrod {M}illman, 56 -- 61.
\newblock \url{https://doi.org/10.25080/Majora-92bf1922-00a}

\bibitem[{Wilcoxon(1947)}]{wilcoxon1947probability}
Wilcoxon, F. 1947, Biometrics, 3, 119.
\newblock \url{https://doi.org/10.2307/3001946}

\bibitem[{Zewdie {et~al.}(2021)Zewdie, Valladares, Cohen, Lary, Ramani, \& Tsidu}]{zewdie2021data}
Zewdie, G.~K., Valladares, C., Cohen, M.~B., {et~al.} 2021, Space Weather, 19.
\newblock \url{https://doi.org/10.1029/2020SW002639}

\bibitem[{Zhao {et~al.}(2020)Zhao, Zank, Adhikari, Hu, Kasper, Bale, Korreck, Case, Stevens, Bonnell, de~Wit, Goetz, Harvey, MacDowall, Malaspina, Pulupa, Larson, Livi, Whittlesey, \& Klein}]{zhao2020identification}
Zhao, L.-L., Zank, G.~P., Adhikari, L., {et~al.} 2020, The Astrophysical Journal Supplement Series, 246, 26.
\newblock \url{https://doi.org/10.3847/1538-4365/ab4ff1}

\bibitem[{Zheng {et~al.}(2017)Zheng, Hu, Chen, \& le~Roux}]{zheng2017automated}
Zheng, J., Hu, Q., Chen, Y., \& le~Roux, J. 2017, Journal of Physics: Conference Series, 900, 012024.
\newblock \url{https://doi.org/10.1088/1742-6596/900/1/012024}

\end{thebibliography}
\bibliographystyle{aasjournal}



\end{document}